\documentclass{article}

\usepackage{arxiv}

\usepackage[utf8]{inputenc} % allow utf-8 input
\usepackage[T1]{fontenc}    % use 8-bit T1 fonts
\usepackage{hyperref}       % hyperlinks
\usepackage{url}            % simple URL typesetting
\usepackage{booktabs}       % professional-quality tables
\usepackage{amsfonts}       % blackboard math symbols
\usepackage{nicefrac}       % compact symbols for 1/2, etc.
\usepackage{microtype}      % microtypography
\usepackage{lipsum}		% Can be removed after putting your text content
\usepackage{graphicx}
\usepackage[numbers]{natbib}
\usepackage{doi}

\title{An Innovative Solution: AI-Based Digital Screen-Integrated Tables for Educational Settings}

\author{ \href{https://orcid.org/0009-0007-8038-1278}{\includegraphics[scale=0.06]{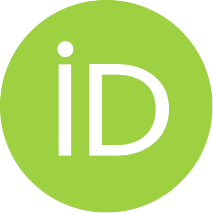}\hspace{1mm}Sagar Tamang}\thanks{
    Correspondance can be addressed to \textit{cs22bcagn033@kazirangauniversity.in}} \\
	Department of IT\\
	The Assam Kaziranga University\\
	Jorhat, India \\
	\texttt{cs22bcagn033@kazirangauniversity.in} \\
	\And
	\href{https://orcid.org/0000-0001-7809-5220}{\includegraphics[scale=0.06]{orcid.pdf}\hspace{1mm}Dr. Dibya Jyoti Bora} \\
	Department of IT\\
	The Assam Kaziranga University\\
	Jorhat, India \\
	\texttt{dibyajyotibora@kazirangauniversity.in} \\
}

\renewcommand{\shorttitle}{AI-Based Digital Screen-Integrated Tables for Educational Settings}

\hypersetup{
pdftitle={An Innovative Solution: AI-Based Digital Screen-Integrated Tables for Educational Settings},
pdfsubject={cs.AI},
pdfauthor={S. Tamang, D. J. Bora},
pdfkeywords={digital, screen, table, classroom, university},
}

\begin{document}
\maketitle

\begin{abstract}
In this paper, we have gone through different AI-Based frameworks used for various educational tasks like digital customized assignment allotment and performance monitoring, identifying slow-learners and fast-learners, etc. application describes a novel invention, digital screen-integrated tables, designed specifically for educational settings. The tables feature integrated digital screens controlled by a central processing unit (CPU), enabling synchronized display of educational content such as textbooks, presentations, exam questions, and interactive learning materials. Additionally, the invention facilitates the collection of student performance data during classroom activities and assessments. The gathered data is utilized for analysis using machine learning models to identify patterns and trends in student learning behaviours. By leveraging machine learning algorithms, educators can ascertain whether a student is a fast learner or a slow learner, based on which, the teacher can allocate more resources to the slow learners. This innovative approach aims to address the evolving needs of modern classrooms by providing a dynamic and data-driven learning environment. The unique integration of digital screens into traditional classroom furniture represents a significant advancement in educational technology. This patent filing encompasses the design, functionality, and method of operation of the digital screen-integrated tables, emphasizing their innovative features and applications in educational institutions.
\end{abstract}

% keywords can be removed
\keywords{digital \and screen \and table \and classroom \and university}

\section{Introduction}
\label{sec:introduction}
Children today are “native speakers” of the digital language of computers, i.e., video games and the internet \cite{Ritthipruek2024}. As such the case, the teachers of the 21st Century shall teach in a way that embraces the changes to better meet the needs of the digital learners \cite{Ritthipruek2024,Prensky2005}. The usage of technology in school environments is very rare, and little change has been observed for the same \cite{AgostiniBiase21}. Adding to the fact that knowledge flow in the classrooms is largely uni-directional and non-participative in nature, even though the research suggests that participative learning, i.e., one where there is cooperation and social learning, is far more effective \cite{AgostiniBiase21}. 

There have been several works in the past that aim to develop environment-embedded networked digital devices in educational settings, also referred to as “ubiquitous computing” \cite{Weiser1993}. One such particular example is the NIMIS project \cite{Hoppe2000}. Such technology-integrated educational settings generate data that can further be used \cite{AgostiniBiase21,Apicella2022}.

\section{Background}
\label{sec:background}
There have been many works done in the past to identify cognitive development in classrooms but not enough on the usage of Machine Learning (ML) algorithms to identify the same \cite{Apicella2022,Helme2001}. However, there have been many works done in the past that integrate technology in educational settings \cite{Daoud2020}. According to international surveys, it was observed that new government policies in both developed as well as developing countries have given a push for the integration of technologies in the classroom \cite{Daoud2020}. These digital devices are of a wide range, such as laptops, tablets, smartphones, whiteboards, etcetera \cite{Daoud2020}. These advancements have now allowed for better data collection, necessary for complex Machine Learning algorithms \cite{Baruah2022,tamang2024performanceevaluationtokenizerslarge}. But, very little has been done to apply the machine learning algorithms to gauge the performance of the students in the classrooms\cite{nathtamang}.

To address this gap, we propose a technology-integrated, digital screen-integrated table to be precise, to foster better collaboration and data retrieval, along with Machine Learning algorithms to identify the slow learners and fast learners in educational settings. 

\section{Description of the Invention}
\label{suc:desc}
In educational settings, students spend 70\% of their time sitting, mostly in classrooms which prompted us to integrate the digital device on the table \cite{Schwenke2022}. As seen in Figure 1, the invention proposes all the individual tables have their processing power, which is integrated into a central node i.e., the teacher. 

\begin{figure}
    \centering
    \includegraphics[width=0.8\linewidth]{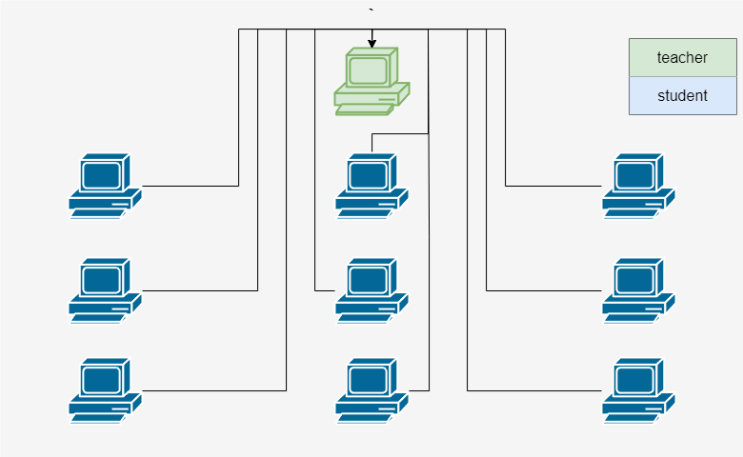}
    \caption{Network of Computers in a Classroom Environment}
    \label{fig:fig1}
\end{figure}

As in the case of the NIMIS project, we would also be developing personalized software along with the hardware that ensures smooth data collection through frequent tests, hence promoting collaboration \cite{Hoppe2000}. In this patent application, we will be utilizing the data gathered from the classrooms to further apply the machine learning algorithms.

\section{Applications}
\label{sec:app}
The digital screen-integrated tables are not limited to taking tests, but for other various tasks as well, promoting collaboration in educational settings. For example, the course materials can be displayed from the screens, live transcription can be generated for person-with-disabilities, etcetera.

\section{Advantages and Benefits}
\label{sec:adv}
In educational settings, most of the time students spend their time sitting down at a table \cite{Schwenke2022}, so this invention targets where the time is spent. This invention also deals with the sluggish adoption of technology in education settings \cite{AgostiniBiase21}. This invention promotes rapid data collection which is necessary to apply machine learning applications \cite{Baruah2022}. The collected data will be used to continuously gauge the performance of the slow learners and fast learners, helping the needy students to get their much-required attention from the teachers, theoretically making the educational settings more effective.

\section{Patent Claims}
\label{sec:patent-claims}
A system for educational settings comprising a display integrated into a table, wherein said display is capable of presenting educational as well as other multi-media content to users.

The system of claim 1, wherein multiple tables are connected together in a network to facilitate collaborative learning experiences in educational settings.

The system of claim 1, further comprising software installed on the display-integrated table for gathering data from users at frequent intervals.

The system of claim 3, wherein said gathered data is processed using machine learning algorithms to assess user performance and provide insights for decision-making by educators.

\section{Summary of the Invention}
\label{sec:summary}
In this patent application, the innovative solution of AI-Based Digital Screen-Integrated Tables for Educational Settings has been described in detail. The invention addresses for the need of a data-driven learning experiences for the classroom of 21st century, wherein traditional furniture are integrated with a digital screen or a synchronized display. The innovation promises to promote collaboration, efficient data collection, and sound decision making. The key features of the innovation includes every tables to have their own display for individual students, regular collection of data from the students, and utilization of machine learning algorithms to identify slow learners and fast learners. The unique design and functionality of the digital screen-integrated table represents a significant advancement in educational technology, offering benefits for students, teachers, and educational institutions.

\section{Acknowledgment}
The authors would like to thank The Assam Kaziranga University for sponsoring this research patent. 

% \bibliographystyle{unsrtnat}
% \bibliography{references}  %%% Uncomment this line and comment out the ``thebibliography'' section below to use the external .bib file (using bibtex) .


\begin{thebibliography}{9}
\bibitem{Ritthipruek2024}
J. Ritthipruek, “All On Screen: The Effects of Digitized Learning Activities on Increasing Learner Interest and Engagement in EFL Classroom.” [Online]. Available: www.iafor.org.

\bibitem{Prensky2005}
M. Prensky, “Teaching Digital Natives: Partnering for Real Learning,” 2005. Accessed: Mar. 17, 2024. [Online]. Available: https://marcprensky.com/wp-content/uploads/2013/04/Prensky-TEACHING\_DIGITAL\_NATIVES-Introduction1.pdf.

\bibitem{AgostiniBiase21}
A. Agostini and E. Di Biase, “Large multi-touch screens to enhance collaboration in the classroom of the 21st century: an Italian experiment.”

\bibitem{Weiser1993}
M. Weiser, “Some computer science issues in ubiquitous computing,” Commun. ACM, vol. 36, no. 7, pp. 75–84, Jul. 1993, doi: 10.1145/159544.159617.

\bibitem{Hoppe2000}
U. Hoppe, A. Lingnau, I. Machado, A. Paiva, R. Prada, and F. Tewissen, “Supporting Collaborative Activities in Computer Integrated Classrooms-the NIMIS Approach,” IEEE CS Press, 2000.

\bibitem{Apicella2022}
A. Apicella, P. Arpaia, M. Frosolone, G. Improta, N. Moccaldi, and A. Pollastro, “EEG-based measurement system for monitoring student engagement in learning 4.0,” Sci Rep, vol. 12, no. 1, Dec. 2022, doi: 10.1038/s41598-022-09578-y.

\bibitem{Helme2001}
S. Helme and D. Clarke, “Identifying cognitive engagement in the mathematics classroom,” Mathematics Education Research Journal, vol. 13, no. 2, pp. 133–153, 2001, doi: 10.1007/BF03217103.

\bibitem{Daoud2020}
R. Daoud, “Using Digital Devices in Classroom Learning: A Complexity Theory Perspective.” [Online]. Available: https://www.researchgate.net/publication/347390796.

\bibitem{Baruah2022}
A. Baruah, J. Goswami, D. Bora, and S. Baruah, “A Comparative Research of Different Classification Algorithms,” 2022, pp. 631–646. doi: 10.1007/978-981-16-2422-3\_50.

\bibitem{Schwenke2022}
P. Schwenke and M. Coenen, “Influence of Sit-Stand Tables in Classrooms on Children’s Sedentary Behavior and Teacher’s Acceptance and Feasibility: A Mixed-Methods Study,” Int J Environ Res Public Health, vol. 19, no. 11, Jun. 2022, doi: 10.3390/ijerph19116727.

\bibitem{tamang2024performanceevaluationtokenizerslarge}
Sagar Tamang and Dibya Jyoti Bora.
\newblock Performance Evaluation of Tokenizers in Large Language Models for the Assamese Language.
\newblock arXiv preprint arXiv:2410.03718, 2024.
\newblock Available at: \url{https://arxiv.org/abs/2410.03718}.

\bibitem{nathtamang}
B. Nath, S. Tamang, S. Munshi, K. K. Pandey, S. Kumar, and P. Randhawa,
\newblock "A Comparative Study of Model Variations: English-Nepali Language Pair,"
\newblock in \emph{2024 OPJU International Technology Conference (OTCON) on Smart Computing for Innovation and Advancement in Industry 4.0}, Raigarh, India, 2024, pp. 1-6.
\newblock doi: 10.1109/OTCON60325.2024.10687932.

\end{thebibliography}

\end{document}